\documentclass[natbib209,12pt,preprint]{aastex}

\newcommand{\msun}{M$_\odot$}
\newcommand{\rsun}{R$_\odot$}
\newcommand{\lsun}{L$_\odot$}

\newcommand{\ace}{$\alpha$}

\shorttitle{The \ace\ formalism}
\shortauthors{De~Marco et al.}

\begin{document}
\title{The role of planets in shaping planetary nebulae}


\author{Orsola De Marco \& Noam Soker}
\altaffiltext{1} {Department of Physics, Technion$-$Israel
Institute of Technology, Haifa 32000 Israel;
soker@physics.technion.ac.il.}

\begin{abstract}
In 1997 Soker laid out a framework for understanding the formation and shaping of planetary nebulae (PN). Starting from the assumption that non-spherical PN cannot be formed by single stars, he linked PN morphologies to the binary mechanisms that may have formed them, basing these connections almost entirely on observational arguments. In light of the last decade of discovery in the field of PN, we revise this framework, which, although simplistic, can still serve as a benchmark against which to test theories of PN origin and shaping. 

Within the framework, we revisit the role of planets in shaping PN. Soker invoked a planetary role in shaping PN because there are not enough close binaries to shape the large fraction of non-spherical PN. In this paper we adopt a model whereby only $\sim$20\% of  all $1-8$~\msun\ stars make a PN. This reduces the need for planetary shaping. Through a propagation of percentages argument, and starting from the assumption that planets can only shape mildly elliptical PN, we conclude, like in Soker, that $\sim$20\% of all PN were shaped via planetary and other substellar interactions but we add that this corresponds to only $\sim$5\%  of all $1-8$~\msun\ stars. This may be in line with findings of planets around main sequence stars. PN shaping by planets is made plausible by the recent discovery of planets that have survived interactions with red giant branch (RGB) stars.

Finally, we conclude that of the $\sim$80\% of $1-8$~\msun\ stars that do not make a PN, about one quarter do not even ascend the AGB due to interactions with stellar and substellar companions, while three quarters ascend the AGB but do not make a PN. Once these stars leave the AGB they evolve normally and can be confused with post-RGB, extreme horizontal branch stars. We propose tests to identify them.


\end{abstract}

\keywords{planetary nebulae: general --- stars: AGB and post-AGB --- stars: mass loss}

\section{Introduction}
\label{sec:introduction}
\citet{Soker1997} established an observationally-based framework whereby planetary nebulae (PN) with different morphologies were mapped to the shaping mechanism that could produce them. These mechanisms ranged from single stars to a multitude of binary interactions with stellar and substellar (brown dwarf and planetary) companions. 

Many discoveries, both observational and theoretical have been made in the last ten years that have thickened the debate of what shapes PN. Key to the debate have been the theoretical work of \citet{Soker2006} and \citet{Nordhaus2007} (but see previous work by \citet{Soker2002b}), who described the difficulty with which single asymptotic giant branch (AGB) stars can sustain rotation and global magnetic fields for long enough to affect the geometry of the mass-loss and the shape of the subsequent PN. In the case of {\it single stars}, rotation and global magnetic fields, have been the leading mechanism suggested for the shaping of non-spherical PN \citep[e.g.,][]{GarciaSegura2005}. Lacking a full understanding of how single AGB stars can produce winds that greatly diverge from a spherical distribution, several authors started looking more favourably to binary origin explanations (for a review see \citealt{DeMarco2009b}), where the companion is either a AGB wind shaping agent (i.e., via gravity), or it induces primary envelope rotation and magnetic fields.

Irrespective of what camp of the debate each researcher favours, the scheme of \citet{Soker1997} has been used as a basis against which to test various hypotheses \citep[e.g.,][]{GarciaSegura1999,Parker2006,Lu2009}. Therefore, in light of the last decade of observational and theoretical results, there is scope for a revision of that framework. In particular, the role of planets in shaping PN was already discussed in the late 90s, but could not be observationally quantified at that time. Today, we know more on the statistics of planetary systems, and, although many questions still remain, this is enough to update our estimate of the role of planets in shaping PN. 

Finally, new planetary discoveries are changing our understanding of stars. New questions are being asked as to the influence planets have on stellar evolution, and, conversely, how stars affect planet evolution and survival (see for instance the proceedings of the conference {\it Planets Beyond the Mains Sequence}; Bamberg, Germany, August 11-13, 2010). The role of planets in PN shaping is topical in this context.

In \S~\ref{sec:Soker1997} we explain the PN shaping framework laid out by \citet{Soker1997}. In \S~\ref{sec:newresults} we discuss new results in the field of PN (\S~\ref{ssec:PNresearch}) and planets (\S~\ref{ssec:exoplanetsresearch}), which allow us to reassess the work of \citet{Soker1997} in \S~\ref{sec:reframingSoker1997}. In \S~\ref{sec:nakedCS} we present some arguments regarding the detection of naked central stars, while in \S~\ref{sec:Summary} we conclude.

\section{The PN shaping framework in 1997}
\label{sec:Soker1997}

Soker (1997) presented a classification of 458 PN, based on the morphological systems of \citet{Schwarz1993} and \citet{Corradi1995}, with the aim of distinguishing between the role of stellar and substellar companions. \citet{Soker1997} further classified elliptical PN into those with large and small departure from sphericity. He divided PN morphologies into four categories and flanked each category with the physical process most likely to give rise to that shape using a series of observational arguments.   
\newline
(1) Spherical PN are formed by progenitors that did not have a companion, or did not interact with their companion.
Some departure from sphericity (though no axi-symmetry) may be expected if the companion separation is large, but still interacting in some way with the primary and the PN is formed over a span of time comparable with the orbital period.  
\newline  
(2) Bipolar PN are formed by a close stellar companion that avoided a common envelope phase, 
or that entered the common envelope only late in the evolution. 
The stellar companion accretes a substantial fraction of the AGB wind, and blows
two opposite jets that shape the nebula into a bipolar structure. Point symmetric structures are
also possible due to precession of the accretion disk around the companion.  
\newline
(3) Elliptical PN with large departure from sphericity are formed by stellar companions in a common envelope.
The companion in the envelope does not blow 
strong jets (or blows no jets at all), but the interaction with the envelope ensures high equatorial mass loss 
that leads to an expanding ring structure.  
\newline
(4) Elliptical PN with small departure from sphericity are formed by a substellar companion (a brown dwarf or a planet)  that spins-up the envelope of the AGB progenitor. Weak jets may 
be present. 

The framework above was based primarily on a series of key observational results and in part on a handful of theoretical studies, which we summarise below:

\begin{itemize} 
\item The theoretical arguments was then, and is still now, that the axi-symmetric structure
(including point-symmetric) {\it requires} the presence of a stellar or substellar companion \citep{Soker2001b,Soker2004,Soker2006}. The rarer point-symmetric PN cannot be explained by
a model based on shaping by the fast wind during the PN phase, and seem to
require precessing accretion disks.  Theoretical models showed that binary interactions have the ability to produce the observed PN morphologies \citep[e.g.,][]{Morris1987}. 

\item From population synthesis studies the fraction of PN that were derived from binary interaction was far lower than the needed 90\% (that is the entire PN population minus 10\% spherical PN). For instance, \citet{Han1995} derived that only $38\pm4$\%, of all PN derive from binary interactions (common envelopes, mergers and wider binary interactions).

\item We knew observationally that $\sim 10\%$
of all PN are round. We also knew that $\sim 15$\% of PN are bipolar while the remaining $\sim$75\% are elliptical (mildly, extremely, and  with or without additional structures).  The detailed classification does change from author to author \citep[e.g.,][]{Corradi1995} so we refrain here from describing additional subdivisions.

\item At the time it appeared that the PN observed around post-common envelope central stars were not  bipolar, but rather elliptical \citep{Soker1997}.

\item Bipolar PN have relatively higher expansion velocities, more in line with the escape velocities of main sequence rather than AGB stars \citep{Corradi1995}.  Elliptical PN, on the the hand, have expansion velocities that can be explained 
by winds from AGB stars.  

\item The spherical halos of many PN, are much fainter than the
elliptical inner region. In the binary model this is accounted for by a binary interaction 
that occurs at a late stage in the superwind phase, the period of enhanced mass-loss that is observed to take place towards the end of the AGB \citep{Delfosse1997}.  The interaction both increases the mass-loss rate and causes the departure from axisymmetry; even a planet can play a role in this process \citep{Soker2000}.

\item Planet statistics were poor in that year.

\end{itemize}

Later, Soker (2001a), realising the possible importance of wider stellar companions, added a fifth class of wide (but not very wide) companions that accrete mass and blow jets. The shaping is only by the jets. A moderately elliptical PN is formed, with pronounced signature of jets (although not as strong as in scenario 2 above). Finally, \citet{DeMarco2005} and \citet{Soker2005} introduced the concept that not all $1-8$~\msun\ stars actually make a PN. Single stars can make a spherical PN of a sub-luminous nature that is harder to detect; \citet{Soker2005} termed this class ``hidden PN".

In Table~\ref{tab:binary} we summarise the conclusions of Soker (1997; column 3), Soker (2001a,b; column 4) and Soker \& Subag (2005; column 5), based on the results above, as well as the conclusions of the present work, that will be justified in the following sections. Once the spherical PN are accounted for by mechanisms involving no interaction and bipolar PN are accounted for by  close companions outside the common envelope ($\sim$10\% and $\sim$15\%, respectively), the remaining $\sim$75\% of all PN needs to be accounted for by other interactions. \citet{Soker1997} expected that the only other PN shaping interaction would be a common envelope interaction, and these interactions are not frequent enough to account for 75\% of all PN. This lead \citet{Soker1997} to conclude that there are not
 enough stellar companions to produce the large fraction of non-spherical PN.  Even including much wider binaries in the class of interacting binaries \citep{Soker2001}, there was still a need for a relatively large number of planetary companions to help shape PN.

\section{In the light of new results}
\label{sec:newresults}

Since the mid-2000s, several results both of a theoretical and observational nature, allowed us to refine the conclusions of Soker (1997,2001a,b) and Soker \& Subag (2005). Here we divide the new results into two classes. Those pertaining PN and those that regard the presence and frequency of planets around main sequence and evolved stars.

\subsection{Highlights of relevant PN research in the last decade}
\label{ssec:PNresearch}

\begin{itemize}

\item The fact that shaping axi-symmetric PN with single stars may be difficult was discussed before \citep[e.g.,][but see also De~Marco (2009)]{Morris1987,Soker1989,Soker1997,Soker2006b,Nordhaus2007}. Recently \citep{Soker2006,Nordhaus2007} it has become even more implausible. Global magnetic fields in AGB stars tend to feed back negatively onto the star's spin, slow it down and shut down the magnetic field itself in less than $\sim$100 years, i.e., shorter than it is needed to shape the superwind $\ga$1000~yr. Magnetic fields in the successful shaping models of, e.g., \citet{GarciaSegura2005} are imposed and constant. The suggestion of \citet{Blackman2001} that tapping part of the convection energy to sustain the magnetic-field forming dynamo is a viable mechanism to sustain AGB fields, was questioned by \citet{Soker2002b} as well as, later on, by the authors themselves \citep{Nordhaus2007}.

\item Several theoretical papers have studied the effects of a binary companion on a mass-losing AGB star. These papers show that the relation between the binary interaction type and resulting PN morphology may not be as straight forward as envisaged by the simple scheme of \citet{Soker1997}: the seminal work of \citet{Mastrodemos1999} studied the effects of a companion accreting from an AGB star wind and forming a stable accretion disk and jets, \citet{ReyesRuiz1999} and \citet{Nordhaus2006} studied the formation of accretion disks and jets around the primary's core by disrupted companion. Further, \citet{Blackman2001} concluded that more massive companions may blow powerful jets that can inflate bipolar lobes. Equatorial flows can be formed by companions in common envelope interactions \citep{Sandquist1998} and in wider orbits \citep{Edgar2008}, but it is not clear whether bipolar morphologies can arise from these systems \citep{Soker1997}. Finally the PN morphology may change after it is initially established at the hand of the fast post-AGB wind \citep{HuarteEspinosa2010}, although it is not known exactly which types of structures are susceptible to this change and which are not.

\item Population synthesis simulations that predict the fraction of PN shaped by binary interactions \citep[e.g.,][]{Han1995}, start from the assumption that {\it all} stars in the main sequence mass range $1-8$~\msun\ will form a PN. Using a different population synthesis technique, \citet{Moe2006} predicted the number of PN (with radius $<$ 0.9 pc) present today in the Galaxy, if all  of the $1-8$~\msun\ stars actually make a PN (46\,000$\pm$13\,000 objects). This number was recently updated by \citet{Moe2010} to 61\,000$\pm$17\,000 objects. They compared this prediction with the actual number of observed PN (with radius $<$0.9~pc - this radius limit is key to avoid much larger uncertainties that would derive from PN detection) of 8000$\pm$2000, derived from extragalactic PN counts \citep{Jacoby1980}, or 11\,000$\pm$2000, derived from a local sample count \citep{Frew2008b}. The predicted number is discrepant with the observationally-based estimate of the Galactic PN population at the 3$\sigma$ level. This calculation supports the earlier suggestion of \citet{Soker2005} that not all stars in the $1-8$~\msun\ range actually make a PN. {\it If so, the PN population does not need to mirror the main sequence population, but it derives from only a subset of it.} In Table~\ref{tab:binary} we reserved a row to ``naked central stars". We exchange the term {\it hidden PN} used by Soker \& Subag (2005) with {\it naked central star of PN}, to reflect the observational consequence that these will not be classified as a PN at all. Still, we do predict that very deep observation may reveal very intrinsically faint PN that are expected to be spherical. While Soker \& Subag (2005) listed together the spherical PN, whether observed or hidden, in Table 1 we classified them into two groups: spherical PN and naked central stars.

\item \citet{Zijlstra2001}, \citet{DeMarco2009b} and \citet{Miszalski2009b} catalogued the morphologies of PN associated with post-common envelope binaries, concluding that there is a preferential association between these binaries and bipolar PN (although not all post-common envelope central stars do have a bipolar PN). This releases the constraint of Soker (1997) to associate bipolar PN with binary systems that avoided a common envelope interaction. In recent years  the issue of what conditions dictate that a binary avoids a common envelope interaction has been hotly debated \citep[e.g.,][]{Beer2007,Carlberg2009,Bear2010,Nordhaus2010}, including when a binary enters a common envelope at a late stage \citep{Corradi2011}. Therefore, at this point we consider all close binary interactions (common envelope and not) in the same group.

\item The MASH survey \citep{Parker2006,Parker2006b,Miszalski2008} revised the fraction of spherical PN from $\sim$10\% to $\sim$19\%, where the fraction of spherical PN is higher for fainter/older PN. Decreasing the fraction of non-spherical PN to $\sim$81\%, this result releases slightly the pressure on the needed number of binary interactions. The MASH survey also slightly decreased the fraction of bipolar PN to 13\%\footnote{We note here that of all the MASH PN 14\% are either stellar (unresolved), irregular or asymmetric. The remaining PN are ellipticals (54\%), round (19\%) or bipolar (13\%). In this paper we absorb the 14\% of stellar, irregular or asymmetric PN in the elliptical class, such that elliptical, round and bipolar PN comprise 68\%, 19\% and 13\%, respectively. Clearly this is likely to artificially inflate the elliptical class, but since this is the most numerous class the percentage error brought by this inclusion remains relatively small.}.

 \item Further, if very wide stellar companions shape mass-loss by accreting AGB wind and blowing jets, the number of possible binary interactions that can create non spherical PN is greatly increased. An example of such wide interactions is Mira AB, where despite the large orbital separation of $\sim 40$ times the AGB stellar radius \citep[338-402~\rsun;][]{Woodruff2004}, the companion influences the wind morphology in the region between the two stars \citep{Marengo2001,Karovska1997,Karovska2005,Matthews2006,Karovska2006}.

\item  \citet{Miszalski2009} revised the fraction of PN binaries with period shorter than $\sim$15 days to 17$\pm$5\% (only slightly larger than the 10-15\% determined by \citet{Bond2000}).  This fraction is a lower limit, due to the photometric precision of the survey used, and the various survey biases (such as the angle between the orbital axis and the line of sight). However, accounting for such biases would not increase this fraction dramatically: \citet{Bond2000} had assumed that the period distribution of the detected systems (P$<$3~days) meant that longer periods could not be observed and so a large fraction of binary PN may be hiding at periods slightly longer than 3~days. On the other hand \citet{Miszalski2009} determined observationally and \citet{DeMarco2008c} determined theoretically that the detection limit of this survey technique is $\sim$2 weeks and that the paucity of central star binaries in the period range 3 days to 2 weeks is real. We can therefore state that while the fraction of post common envelope binaries is only slightly larger than 17\%, the fraction of central stars in binaries with periods longer than 2 weeks remains unknown and could be large.

\item We still do not know the fraction of PN which have binary central stars in the period range 2 weeks to the maximum period that still allows for an interaction. Preliminary work by \citet[][see also Passy et al. (2011, in preparation)]{DeMarco2010b} indicates that this fraction may be larger than predicted by the current PN scenario where almost all 1-to-8~\msun\ stars make a PN. If confirmed, this finding is in line with the conclusions of \citet{Soker2005} and \citet{Moe2006}.


\end{itemize}

\subsection{Highlights of relevant exoplanet research in the last decade}
\label{ssec:exoplanetsresearch}

About a decade ago estimates of the percentage of planet-hosting solar-like stars stood at $\sim 3 \%$, although it was already suspected that the percentage might be much higher for stars with higher metallicities: $\sim 25\%-30\%$ for stars with twice the solar metallicity \citep[$\rm{Fe/H}>0.3$; e.g,][]{Santos2004,Fischer2005}. \citet{Lineweaver2003} extrapolated from the detected parameter space of planetary systems to below detection sensitivity and concluded that the real fraction of planet hosting Sun-like stars should be at least 9\% for $M_p \sin i > 0.3M_{\rm J}$ and $P < 13$ years and at least 22\% for  $M_p \sin i > 0.1M_{\rm J}$ and $P < 60$~yrs. They also suggested that since this area of the planet mass-period plane is only a fraction of that occupied by our own Solar System, these fraction may be even larger once the entire parameter space is sampled. This was the situation when Soker \& Subag (2005) updated the PN formation channel statistics (See Column 5 of Table~1).

Today we have strong evidence that the planetary fraction increases with both metallicity and mass of the host star \citep{Johnson2010}, although our knowledge of the planet-hosting star fraction as a function of planet mass and orbital separation is grossly incomplete. A new finding in this respect is that of \citet{Bowler2010} who determined that $\sim  26^{+9}_{-8}\%$ of stars having a main sequence mass of $1.5 \la M/$\msun$ \la 2.0$ host massive planets at {\it large separations} (but still less than 3~AU). The new finding is not only of a larger fraction of planet-hosting stars, but also puts the planets at larger orbital separations, where they can more easily be available to shape winds from AGB stars: if the planets are too close to their host, they interact with the star during the RGB phase, and either prevent the star from reaching the AGB at all (and form a PN), or they just get destroyed before the star reaches the AGB \citep{Nelemans1998,Villaver2007b,Villaver2009}. There is also an indication from the analysis of a handful of thick disk stars that the planetary fraction for low metallicity, {\it old stars} may be higher than for younger stars at similar metallicities \citep{Sheehan2010}.

Another relevant finding is that of brown dwarf and planetary mass companions around extreme horizontal branch (EHB) stars, also termed sdO and sdB subdwarfs. Although such findings regard stars that have gone through the RGB, {\it not} the AGB evolution, we can draw information of a general interest. EHB stars are on the horizontal branch (HB) and have small envelope mass, because their RGB progenitor lost most of its envelope \citep{D'Cruz1996}. Current consensus is that in most cases the mass loss was caused or enhanced by the interaction with a stellar-mass companion \citep{Han2002}. 

\begin{itemize}

\item Companions with likely brown dwarf or planetary masses have been discovered at 1 to a few AU from single \citep{Silvotti2007} or  close binary \citep{Lee2009, Beuermann2011, Qian2009b, Qian2009} EHB stars. For the latter group, the companions orbit close binaries that went thought a common envelope interaction when the RGB progenitor of the sdB star engulfed its stellar companion. It is not clear where these tertiary/low mass companions were located within the pre-common envelope binary, but they may have survived despite the dynamical mayhem of the common envelope interaction\footnote{It is not excluded that these planets formed in the ejected common envelope \citep{Perets2010}.}.

\item \citet{Geier2009} discovered a companion with mass $8-23 M_J$ (either a planet
or a low mass brown dwarf) with period of only 2.391~days around HD149382.
This is the first detection of a substellar object, with a lower possible mass reaching the planet domain, that definitely went through a common envelope evolution inside an RGB envelope. The survival of the planet implies that interactions between gas giants and stars alter stellar evolution, therefore if a planet at a suitable distance is present around a growing AGB star, a common envelope with such planet may be survived and lead to shaping of the ejecta. A second possible planet ($M \sin i = 1.25$~M$_{\rm J}$) was recently discovered by \citet{Setiawan2010} at 0.116~AU from the metal poor HB star HIP~13044.
\end{itemize}



\section{Updating the PN shaping framework}
\label{sec:reframingSoker1997}

In light of the new discoveries ({\S~\ref{sec:newresults}), we reframe here the results of \citet{Soker1997} and subsequent updates by \citet{Soker2001,Soker2001b} and \citet{Soker2005}.  The work discussed in \S~\ref{ssec:PNresearch} releases the tight constraint that brought \citet{Soker1997} to conclude that a large number of PN should be shaped by planets.  The increasing numbers of brown dowarf and planet discoveries around evolved stars also give support to the idea that planets play a role in shaping PN, provided that they are at a suitable distance from the star to interact during the AGB. Finally, the larger fraction of planets detected recently around main sequence stars, compared to what was believed a decade ago, provides us with more flexibility when discussing the role of planets in shaping PN.

In what follows we quote percentage figures  with an accuracy of 1\%. This accuracy is unrealistically high. However, we do so for ease of following some of the arguments. At the end, we will discuss what should be considered reasonable errors on these percentage estimates.

We start by returning to the discussion of \S~\ref{ssec:PNresearch}: instead of considering the PN population as the inevitable child of the $1-8$~\msun\ star population, needing to reflect its binary fraction, we can think of the PN population as deriving from only a {\it subset} of the $1-8$~\msun\ stars. 
We adopt the subdivisions of PN morphological types envisaged by \citet{Soker1997}, with the changes brought in by the MASH morphological discoveries (\S~\ref{sec:Soker1997}; Table~\ref{tab:morphologies}). 
 
First of all, we determine the fraction of all $1-8$~\msun\ stars that result in a PN {\it of any shape}. \citet{Moe2010} concluded that only 18$\pm$8\% of all $1-8$~\msun\ stars (11\,000/61\,000) make PN and 82$\pm$8\% make no PN, or do not go through the AGB at all (see \S~\ref{ssec:PNresearch}).  This estimate rests on single stellar evolution, the initial mass function and the galactic star formation history, as well as observations. It is almost independent of the binary fraction and period distribution.

An independent way to determine the fraction of all $1-8$~\msun\ stars that result in a PN of any shape is to consider only the main sequence binary fraction and period distribution. A summary of how these fractions are determined can be found in Table~\ref{tab:AllStars}. Approximately 57\% of F and G main sequence stars are in binaries and $\sim$30\% have an orbital separation $\la 30$~AU \citep{Duquennoy1991}, and will interact sometime during the primary star's life. We have taken a maximum separation that is possibly too large for some types of interaction, but is smaller than the distance between Mira and its interacting companion (see \S~\ref{ssec:PNresearch}). We assume, once again, that an interaction is needed to produce non-spherical PN. We also account for the fact that 14\% of all stellar systems suffer a strong interaction on the RGB (orbital separation $\la 3$~AU; \citealt{Duquennoy1991}) that induces so much mass-loss that the system is prevented from ever evolving to the AGB. 
We then deduce that $30-14$=16\% of all stars suffer an AGB interaction with a stellar companion and go on to form a non spherical PN.
This means that $\sim$16\% of all stars go through an AGB interaction, $\sim$14\% go through and RGB interaction and never ascend the AGB and the remaining $\sim$70\% suffer no interaction with a stellar companion and evolve into a naked central stars or a central star with round PN, {\it unless they have an interaction with a substellar companion}. As \citet{Soker1997} we assume that mildly elliptical PN are shaped by interactions with substellar companions. Hence, $\sim$16\% of all stars go through an interaction with a {\it stellar} companion and result in non-round and non-mildly elliptical PN. This is equivalent to stating that $\sim$16\% of all stars result in 61\% of all PN (the non-round and non mildly elliptical ones; Table~\ref{tab:AllStars}), and this means that 26\% (16/61$\times$100) of all $1-8$~\msun\ stars make a PN of any shape. This estimate, which has a probable error of $\pm 5$\%, is based on approximate binary considerations, and is consistent with that outlined above (cf. 26\% with 18$\pm$8\%), obtained by \citet{Moe2010} almost entirely from single stellar evolution arguments and PN counts. 

In Table~\ref{tab:AllStars2} we finally adopt 26\% as the fraction of $1-8$~\msun\ stars that are able to make a PN of {\it any shape}. Approximately 14\% is the fraction of all stars that do not ascend the AGB because of a strong interaction on the RGB. The remaining $\sim$60\% is the fraction of all stars that do not make a PN because they suffer no interaction (for further discussion see \S~\ref{sec:nakedCS}).
Finally, we split 16\% into fractions reflecting the PN morphological types presumed to derive from binary interactions with stellar companions, i.e., 13\% are bipolar (16$\times$0.13 / 0.61=3.4), 28\% are extremely elliptical (16$\times$0.28 / 0.61=7.3), and 20\% are mildly elliptical with jets (16$\times$0.20 / 0.61 =5.2). The remaining 10\% of all 1--8~\msun\ stars (26--16\%) goes to form the two remaining PN classes, the 19\% round PN (10$\times$0.19 / 0.39=4.9) and the 20\% mildly elliptical PN with no jets (10$\times$0.20/0.39=5.1).  

Recently \citet{Moe2010} carried out a more precise binary population synthesis model, completely independent of morphological subdivisions, and determined that 8\% of all $1-8$~\msun\ stars suffer a strong interaction with a stellar companion on the AGB (common envelopes and interactions outside the envelope). This is in line with the approximate estimate based on binary considerations and morphological subtypes, where $\sim$11\% of all $1-8$~\msun\ stars suffer a strong AGB interaction (3.4+7.3\%: Table~\ref{tab:AllStars2} and last column of rows 2 and 3 in Table~\ref{tab:binary}).

In Table~\ref{tab:binary} (last column) we finally list a revised subdivision of the evolutionary channels followed by intermediate mass stars and their link to PN morphology. During the last decade of simulations and observations we have learned that a one-to-one correspondence between morphologies and shaping channels is as improbable as the likelihood of finding two identical PN. This is why we stress (see also note ``c" in Table~\ref{tab:binary}) that these are only guidelines. The take-home messages are that (i) common envelope and other interactions, where the companion is close to the AGB surface, likely result in strong collimation leading to bipolarity and extreme ellipticity; (ii) that low mass companions such as planets are likely only to promote small departures from a spherical shape; (iii) that blowing jets may be promoted not only by an accreting stellar companion at some distance from the primary, but also by a disrupted lower mass companion that forms a disk around the primary core \citep{Nordhaus2006}. If so, this would confuse the distinction between channels 4 and 5 in Table~\ref{tab:binary}.

If, as in \citet{Soker1997,Soker2001} and \citet{Soker2005}, we assume that mildly elliptical PN with no jets were shaped mainly by a substellar companion, we predict that substellar companions have shaped $\sim$20\% of all PN, which corresponds to $\sim$5\% of all $1-8$-\msun\ stars having suffered an interaction with a planetary companion on the AGB.  However, if substellar companions can get disrupted, form a disk around the primary core and blow jets, then this percentages could rise to as much as $\sim$40\% and $\sim$8\%, respectively.

\begin{deluxetable}{cllcccc}
\tabletypesize{\scriptsize}
\tablewidth{0pt} \tablecaption{Evolutionary channels for different morphological types.}
\small
\tablehead{\colhead{} & \colhead{Evolutionary channel} & \colhead{By-product} & & \colhead{Percentage} &&  \\
\colhead{} &\colhead{} &   & \colhead{1997} & \colhead{2001} & \colhead{2005} & \colhead{This work }}
\startdata 1 & No interaction & Spherical   PN& $\sim 10$ & $\sim 10$ & $\sim 10$ & $\sim$5$^a$ (19)$^b$ \\
\hline
2&Close stellar companion  &  &  &   & &  \\
&outside envelope$^c$         & Bipolar PN&$\sim 11$& $\sim 15$& $\sim 15$  & $\sim3$$^a$ (13)$^{b}$ \\
\hline
3&Stellar companion in     & Extremely &  &    & &  \\
&a common envelope$^c$        & elliptical PN& $\sim 23$& $\sim 25$& $\sim 25$ & $\sim 7$$^a$ (28)$^{b}$ \\
\hline
4&Substellar companion   &    Mildly& & & &  \\
&in a common envelope$^d$   &  elliptical PN &$\sim 56$&$\sim 35$& $\sim 15$ & $\sim 5$$^a$ (20)$^b$\\
\hline
5&Wide stellar   & Elliptical PN&$-$ & $\sim 15$& $\sim 15$  & $\sim 5$$^a$ (20)$^b$\\
  &   companion$^d$                    & +jets     &    &          &             &  \\
\hline
6&No interaction & Naked  central &-- &--& $\sim$20&$\sim60^e$\\
   & &    stars (post-AGB) & & & &  \\

\hline
7&Strong interaction  & EHB stars       &   &   &   &   $\sim 14^e$  \\
&on the RGB          &  (post-RGB)    &   &   &   &      \\
\enddata
\tablecomments{Subdivision of the evolutionary channels taken by all $1-8$~\msun\ stars and the evolution of our understanding of the percentages over the past decade. The percentage are from Soker
(1997), \citet{Soker2001,Soker2001b}, \citet{Soker2005} and the present paper.  EHB stars are formed form RGB stars that have lost most of their envelope due to a strong binary interaction with a stellar companion and will never ascend the AGB.}
\tablenotetext{a}{All these percentages should be considered ``a few percent". However, their relative accuracy may be much better than their absolute one. They add up to 25 instead of 26 because of rounding errors.}
\tablenotetext{b}{All numbers in brackets are fractions of the PN population and add up to 100. The distribution of percentages within brackets reflect the distribution of morphological types.}
\tablenotetext{c}{We maintain these two channels separate for historical reasons, although we know that these two evolutionary channels do not always result in the corresponding morphologies. They could be combined into one channel called: strong binary interactions.}
\tablenotetext{d}{It is conceivable that also substellar companions may blow jets if they are destroyed in a common evnelope interaction and form a disk around the primary's core (\S~\ref{ssec:PNresearch} and \citealt{Nordhaus2006}).}
\tablenotetext{e}{These uncertainties are of the order of $\pm$5\%.}
\label{tab:binary}
\end{deluxetable}

\begin{deluxetable}{cll}
\tablewidth{0pt} \tablecaption{Percentages of all PN that reside in different morphological classes.}
\small
\tablehead{ \colhead{Percentage of PN} & \colhead{Class} & \colhead{Reference}}
\startdata 
81\%  & non-round PN  & \citealt{Parker2006}\\
19\%  & round PN & \citealt{Parker2006}\\
13\% & bipolar PN  &\citealt{Parker2006}  \\
28\%  & extremely elliptical PN & \citealt{Parker2006,Soker1997}$^a$\\
20\%  & mildly elliptical PN with no jets& \citealt{Parker2006,Soker1997}$^a$ \\
20\%  & mildly elliptical PN with jets &\citealt{Parker2006,Soker1997}$^a$\\
\enddata
\tablenotetext{a}{Sixty-eight percent elliptical \citep{Parker2006} = 28\% + 20\% + 20\%. The subdivision into three classes follows \citet{Soker1997}. The jet/no jet division is from \citet{Balick1998}.}
\label{tab:morphologies}
\end{deluxetable}

\begin{deluxetable}{clc}
\tablewidth{0pt} \tablecaption{Subdivision of {\it all 1-8}~\msun\ stars into orbital separation groups.}
\small
\tablehead{ \colhead{Percentage } & \colhead{Class}  & \colhead{Method of}   \\
\colhead{ of 1-8~\msun\ stars} & \colhead{}  & \colhead{determination}   }
\startdata 
57\%  & binaries & DM91$^a$ \\
43\%  & single stars& $100-57$\%\\
30\% & binaries with $a<30$~AU -- interaction on the RGB or AGB & DM91$^a$  \\
70\%  & single stars and binaries with $a>$30~AU -- no interaction& $100-30$\%\\
14\%  & binaries with $a<3$~AU --  interaction on the RGB& DM91$^a$ \\
16\%  & binaries with $3<a<30$~AU --  interaction on the AGB& $30-14$\%\\
\enddata
\tablenotetext{a}{\citealt{Duquennoy1991}.}
\label{tab:AllStars}
\end{deluxetable}

\begin{deluxetable}{clc}
\tablewidth{0pt} \tablecaption{Subdivision of {\it all 1-8}~\msun\ stars into binarity and interaction classes.}
\small
\tablehead{ \colhead{Percentage} & \colhead{Class}  & \colhead{Method of }  \\
\colhead{of all 1-8~\msun\ stars} & \colhead{}  & \colhead{determination} }
\startdata 
(14$^a$+$\epsilon^b$)\%  & {\it do not} ascend the AGB and {\it do not} make a PN& DM91; Table~\ref{tab:AllStars}\\
3\%  & make bipolar PN& 16\%$^c$ $\times 0.13 / 0.61^d$\\
7\%  & make extremely elliptical PN& 16\%$^c$ $\times 0.28 / 0.61^d$\\
5\%  & make mildly elliptical PN with jets& 16\%$^c$ $\times 0.20 / 0.61^d$\\
5\%  &make spherical PN& (26-16)\%$^c$ $\times 0.19^d / 0.39$ \\
5\%  & make mildly elliptical PN& (26-16)\%$^c$ $\times 0.20^d / 0.39$\\
26\%  & make PN of {\it any} shape & $\sim(3+7+5+5+5)$\% \\
(60-$\epsilon^b$)\% & ascend the AGB but do not make a PN& $(100-26-14)$\%  \\
\enddata
\tablenotetext{a}{This fraction is tied to the choice of the orbital separation limit (we have selected 3~AU; see Table~\ref{tab:AllStars}) within which binaries and star-planet systems interact on the RGB. This limit is not well known. It also depends sensitively on the primary mass and the strength of the tides.}
\tablenotetext{b}{$\epsilon$ denotes a small fraction of stars that have an interaction with a substellar companion on the RGB and, as a result, does not ascend the AGB. The same small fraction is taken out of the group of stars that do ascend the AGB.}
\tablenotetext{c}{Sixteen percent is the percentage of all stars that have a stellar companion and that suffer an interaction on the AGB, resulting in bipolar, extremely elliptical and mildly elliptical with jets PN; 26\% is the percentage of all stars that make a PN of any shape, so 26-16 is the percentage of all stars that make round and mildly elliptical PN; see Table~\ref{tab:AllStars}.}
\tablenotetext{d}{These are the morphological percentages from Table~\ref{tab:morphologies}.}
\label{tab:AllStars2}
\end{deluxetable}

\subsection{The implied fraction of planets from the AGB perspective}
\label{ssec:planets}

The fact that only $\sim$30\% of main sequence stars have stellar companions close enough
to interact, combined with the fact that $\sim$81\% of PN have non-spherical shapes that appear
to need a companion, lead \citet{Soker1996} to state that
``substellar objects (brown dwarfs or gas-giant planets) are commonly present
within several AU around main-sequence stars.
For a substellar object to have a high probability of being present within
this orbital radius, on average several substellar objects must be present
around most main-sequence stars of masses $ \la 5 M_\odot$.''

We have derived here that $\sim$5\% of all $1-8$~\msun\ stars suffer a strong interaction with a substellar companion {\it on the AGB}. This estimate depends {\it critically} on (i) the {\it assumption} that mildly elliptical PN with no jets are mostly shaped by a substellar companions, (ii) on the frequency ($\sim$20\%) of this specific PN morphology and (iii) on the two independent arguments that predict that that only 18\% or 26\% of $1-8$~\msun\ stars make a PN at all (0.18 or 0.26$\times$0.20 $\sim$ 0.05).

These substellar companions should occupy orbits between $\sim$3 and $\sim$30 AU around the main sequence progenitors. Closer companions (whether stellar or substellar) suffer a strong interaction (likely a common envelope interaction) on the RGB and are either destroyed or preclude the primary from ascending the AGB because of excessive mass-loss \citep{Soker1998b,Bear2010, Nordhaus2010}. Companions farther out\footnote{The limit out to which a companion is
captured tidally during AGB phase, according to \citet{Soker1996}, is $\sim$12~AU. \citet{Nordhaus2010} would revise this limit down. We have however
widened it to 30~AU in light of other types of interactions that do not involve captures. We have cited the AGB star Mira as an example of a wide binary which is nevertheless interacting. Whether this limit applies to low mass companions and planets is not known.} would not interact. 

Using a population synthesis technique one could use this estimate and work backwards to determine the parent population of these star-planet systems. 
Even without such calculation, we can already draw a few conclusions.
The median main sequence progenitor mass of today's PN population is 1.2~\msun\ and the median metallicity is approximately solar \citep{Moe2006}. 
Taking the maximum radii on the AGB, $R_A$,
and on the RGB, $R_R$, from \citet{Iben1985}, \citet{Soker1998} derived the
following approximation for the ratio of the maximum AGB to maximum RGB radius as a function of mass:
\begin{equation}
\log(R_{\rm A}/R_{\rm R}) = 3.7 \log^2 (M/M_\odot) - 0.37 \log (M/M_\odot) + 0.16,
\qquad M \la 2.25 M_\odot,
\end{equation}
for stars which develop degenerate helium cores, and
\begin{eqnarray}
\log(R_{\rm A}/R_{\rm R}) = 2.2 - 1.8 \log (M/M_\odot), \qquad M \ga 2.35 M_\odot,
\end{eqnarray}
for more massive stars, where $M$ is the primary's mass on the zero age main sequence.
From these relations we can state that for stars with spectral type earlier than A the radius ratio is $>2$, while for stars with spectral type later than F-G, which correspond to the progenitors of central stars of PN,  this ratio is $<2$. For the progenitors of PN, therefore, the planetary population has to be relatively far out. We have quoted limits of 3 and 30~AU, respectively. The lower limit depends critically on the adopted prescription of tidal capture and mass-loss \citep{Soker1996,Villaver2007b,Villaver2009,Nordhaus2010}.

Considering that the fraction of solar-like stars hosting relatively close by planets may be of the order of 10-20\% and considering that we predict that a low $\sim$5\% of all solar like stars have substellar companions further out than $\sim$3~AU, it is probable that the substellar companions farther out are the outermost companions in a system of multiple planets.

\section{Do we observe the naked central stars?}
\label{sec:nakedCS}

In this subsection, we want to elaborate further on the suggestion that $\sim$60\% of all $1-8$~\msun\ stars that ascend the AGB do not make a visible PN. This is a surprisingly high number and implies a large population of post-AGB stars which, under the commonly assumed scenario, would have PN, but in the revised scenario do not.  To be more precise, \citet{Moe2006,Moe2010} predict that $\sim$61\,000 stars in the Galaxy are in the evolutionary phase appropriate to be surrounded by a PN of less than 0.9~pc in radius, i.e., these stars are in a post-AGB with $T_{\rm eff} \ga 30\,000$~K, have left the AGB less than 25\,000 years ago, and have mass $\ga 0.55$~\msun. This population is about 6 times larger than the Galactic PN population with radius smaller than 0.9~pc (11\,000 PN; \citealt{Frew2008b}), from which we predict that 50\,000 (61\,000 -- 11\,000) stars in the Galaxy are ``naked" central stars. It would be much harder to detect naked central stars, because of high reddening in the Galactic plane. However, we should consider that the deep MASH survey has now detected $\sim$3000 PN, which is $\la$30\% of the total \citep{Jacoby1980,Moe2006,Frew2008b}. PN are particularly bright and have emission lines in the red part of the spectrum which are less subject to reddening. The fraction of detectable naked central stars should be far smaller.

Naked central stars would look like subdwarf O and B stars, most of which are post-RGB stars, and would be easily confused with them. In some cases they could be told apart from those sdOB stars which are in a post-RGB phase of evolution  on the grounds of higher luminosity, and larger surface gravity (although there is an overlap between post-RGB and post-AGB stars on the $\log g - T_{eff}$ plane \citep[][their Fig.~4]{Napiwotzki1999}). 

The fraction of naked central stars in the sdOB sample should be only a few percent of the total, because of their short lifetimes compared to the post-RGB phase. Any post-AGB object, independent of its mass, will fade below 100~\lsun\ in less than 100,000 years \citep{Vassiliadis1994}. All post-AGB objects with $M>0.60$~\msun\ will fade to $L=100$~\lsun\ in only 10,000 years.  HB lifetimes are instead of the order of several tens of millions of years \citep{Dorman1993}.

We have an indication that the predicted number of post-AGB sdOB stars are there, {\it but this must be corroborated by a more accurate count.} O'Tool (private communication - but see also \citealt{Hirsch2008}), determined that in about 120 sdO stars, there are several objects which {\it could} be in the post-AGB phase. This could constitute the few percent predicted. For the sdB stars, a few hundred of which have determined $\log g$ and $T_{eff}$, several
reside in the $\log g - T_{eff}$ locus appropriate for post-AGB stars \citep{Napiwotzki1999,Hirsch2008}.
There should be fewer sdB post-AGB stars than sdO post-AGB stars, because their evolutionary
times for the temperature $T_{eff} <$ 40\,000~K is quite a lot shorter than for $T_{eff} >$40\,000~K. 

Looking at Fig.~2 of \citet{Napiwotzki1999} or Fig.~3 of \citet{Hirsch2008} we see that there are a number of sdB and sdO stars that are not surrounded by a PN. Several objects are indeed found on the horizontal part of the post-AGB tracks, where a PN is expected.  Others are seen on the descending part, but at high core mass such that one might expect that a PN would be present, considering the shorter evolutionary timescales. This sample is not homogeneous, and it is therefore difficult to determine whether the number of naked central stars on this plot is what is expected from the prediction by \citet{Moe2006,Moe2010}. 

Deep imaging surveyes may reveal a number of faint and spherical PN around naked central stars. The fact that the deep MASH survey approximately doubled the fraction of circular PN may already have bourn out this prediction. 

\section{Summary}
\label{sec:Summary}

The first assumption we adopt in this paper, which is increasingly supported by observations and theoretical considerations, is that single stars are mostly unable to produce PN whose shapes diverge from spherical. Second, we bring to bear the population prediction that the fraction of stars in the initial mass range $1-8$~\msun\ that actually form PN is only $\sim$20\% (see \S~3.1). 

As in the past we make the simplistic assumption that mildly elliptical PN with no jets are formed exclusively by planetary interactions. This assumption is broadly justified by (i) simulations that show that density contrasts drastically reduce for decreasing companion mass in common envelopes \citep{Sandquist1998,DeMarco2003} and outside the common envelope \citep{Kim2010}; such interactions are therefore likely to lead to mildly elliptical shapes at best. (ii) The fact that known PN around close binaries are by and large bipolar or strongly elliptical and have often jets and substructures; if the stellar-mass companions are at larger orbital separations than those that interact in a common envelope, theoretical considerations suggest that accretion onto the companion promotes jets, which, once again, lead to more dramatic departures from spherical symmetry than a mild ellipticity. Despite these justifications, the one-to-one correspondence between interactions with substellar companion and the generation of mildly elliptical PN should be considered only as a guideline to be confronted by observations, and against which to continue testing future shaping theories.


With the premise above, we predict that the fraction of PN shaped by planets ($\sim 20 \%$) corresponds to only $\sim 5$\% of all $1-8$~\msun\ stars having interacted with a planet on the AGB, indicating that a few percent of $1-8$~\msun\ stars should have Jupiter-class companions further out than a few AU, a thing that seems increasingly plausible given new discoveries of planets around main sequence stars \citep[e.g.,][see section 3.2]{Johnson2010,Bowler2010}. Planets farther out may be the outer planets in planetary systems. 

Finally, the implication that $\sim 60 \%$ of all stars in the initial mass range $1-8$~\msun\ do not go through a PN phase, {\it may} be justified observationally. There are several hot subdwarf stars that occupy a location of the $\log g - T_{\rm eff}$ diagram appropriate for central stars of PN, but that do not exhibit a PN. In order to strengthen this claim one would have to make a prediction of the fraction of all subdwarf O and B stars that are expected to be post-AGB in origin through a population synthesis. Also, one would have to homogeneously locate a sufficient number of these stars on the $\log g - T_{\rm eff}$ diagram to determine if the prediction is verified.

\acknowledgements N.S. acknowledge supported by the Asher Fund for Space Research at the Technion, and
the Israel Science foundation. We thank the referee, Adam Frank, for extremely helpful comments.

\bibliographystyle{../../../apj_new}                       
\bibliography{../../../bibliography}

\end{document}